\documentclass[12pt]{article}
\pdfoutput=1
\usepackage{tikz}
\usetikzlibrary{matrix,arrows,decorations.pathmorphing,decorations.pathreplacing}
\usepackage{jheppub}
\usepackage{amsmath,amssymb,euscript,array,mathrsfs,appendix,ctable,marvosym}
\usepackage{mathtools}
\usepackage{arydshln}
\usepackage{todonotes}
\usepackage{graphicx}
\usepackage{color}

\def\BH{{\cal H}}

\newcommand{\Tr}{\operatorname{Tr}}

\def\bra#1{\langle #1|}
\def\ket#1{| #1\rangle}

\def\VEV#1{\langle #1\rangle}

\newcommand{\EQ}[1]{\begin{equation}\begin{split} #1
\end{split}\end{equation}}

\title{Classical from Quantum}
\author{Timothy J. Hollowood}
\affiliation{
Department of Physics, Swansea University,\\ Swansea, SA2 8PP, United Kingdom
}

\emailAdd{t.hollowood@swansea.ac.uk}

\abstract{We consider the quantum-to-classical transition for macroscopic systems coupled to their environments. By applying
Born's rule, we are led to a particular set of quantum trajectories, or an unravelling, that describes the state of the system from the frame of reference of the subsystem. The unravelling involves a branch dependent Schmidt decomposition of the total state vector. The state in the subsystem frame, the conditioned state, is described by a Poisson process that involves a non-linear deterministic effective Schr\"odinger equation interspersed with quantum jumps into orthogonal states. We then consider a system whose classical analogue is a generic chaotic system. Although the state spreads out exponentially over phase space, the state in the frame of the subsystem localizes onto a narrow wave packet that follows the classical trajectory due to Ehrenfest's Theorem. Quantum jumps occur with a rate that is the order of the effective Lyapunov exponent of the classical chaotic system and imply that the wave packet undergoes random kicks described by the classical Langevin equation of Brownian motion. The implication of the analysis is that this theory can explain in detail how classical mechanics arises from quantum mechanics by using only unitary evolution and Born's rule applied to a subsystem.
}

\notoc
\begin{document}

\maketitle

\newpage

\section{Introduction}

Quantum mechanics is a remarkable fusion of a linear, deterministic theory where evolution involves unitary rotations of a vector in a Hilbert space, with stochastic evolution, in the form of Born's rule. The fusion is remarkably successful and predictive: unitary evolution is used to compute an amplitude from which a probability follows via Born's rule. 

Whilst quantum mechanics has unlocked the behaviour of microscopic systems in exquisite detail, can it make the important prediction of a classical world evolving according to Newton's Laws on macroscopic scales? Are new ingredients needed or is the fusion of deterministic and stochastic rules enough to predict that a classical 
world emerges out of the quantum world on macroscopic scales?
\pgfdeclareimage[interpolate=true,width=6cm]{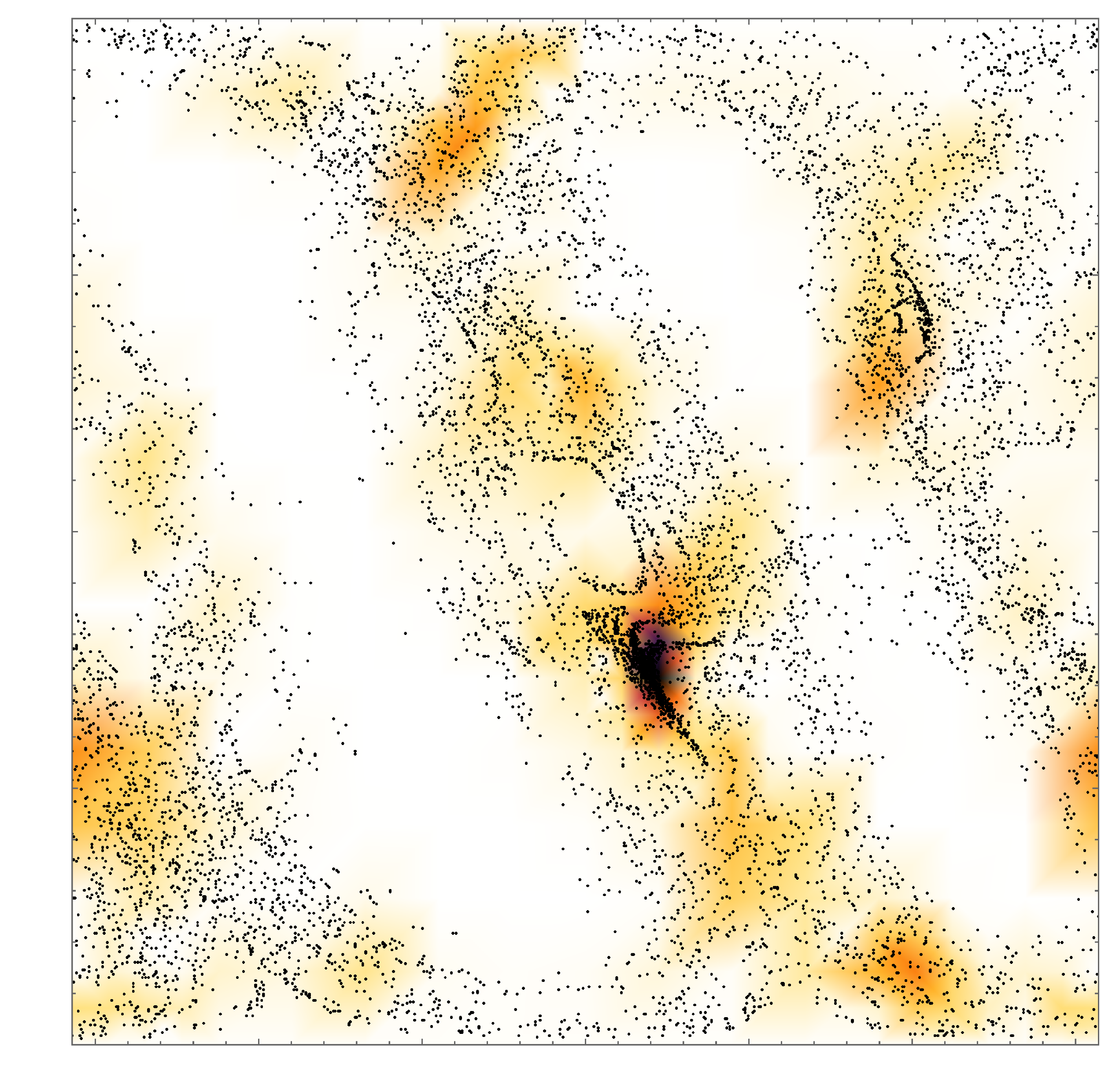}{fig1}
\begin{figure}[!h]
\begin{center}
\begin{tikzpicture}[scale=0.7]
\pgftext[at=\pgfpoint{0cm}{0cm},left,base]{\pgfuseimage{fig1}} 
\end{tikzpicture}
\caption{\footnotesize  For a chaotic kicked rotor, the evolution of a classical ensemble of points and a phase space density plot of the Husimi function of a quantum state. The quantum state begins as a localized coherent state whose probability density is matched by the classical distribution of points. The quantum state follows the classical exponential spreading of the classical points very closely until it fills phase space as shown even though we are far from the correspondence limit: here $J/\hbar=100$. For longer times, the distributions diverge due to quantum interference.}
\label{f6} 
\end{center}
\end{figure}
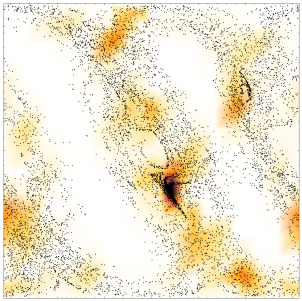

Textbook discussions of the quantum-to-classical transition usually go no further than Ehrenfest's Theorem. The latter applies to states that are narrow wave packets compared with the scale over which the potential $V(x)$ varies. The implication is that the expectation value $\VEV{V'(x)}\approx V'(\VEV{x})$ and so the centre of the wave packet follows a classical trajectory; i.e.~for the expectation values $\bar x=\VEV{x}$ and $\bar p=\VEV{p}$, 
\EQ{
\frac{d\bar x}{dt}=\frac{\bar p}m\ ,\qquad \frac{d\bar p}{dt}=-\VEV{V'(x)}\approx-V'(\bar x)\ .
}
The problem with invoking Ehrenfest's Theorem by itself, is that, even if one could explain the special initial condition, narrow wave packets do not remain narrow. Even wave packets in a free theory spread out, although for a macroscopic system the spreading is very slow. However, the wave function spreads much faster in a system that is classically chaotic. In the classical theory, a small volume in phase space spreads out exponentially in a complicated fractal way to cover a large volume (while maintaining its true volume in accord with Liouville's Theorem). The exponential spreading is defined by the effective Lyapunov exponent $\lambda$, so the separation between two trajectories diverges as $\exp[\lambda t]$. In the quantum theory this exponential spreading is mirrored by the quantum state: see figure \ref{f6} for an example of a chaotic kicked rotor. In a phase space picture, provided by the Wigner or Husimi function, a coherent state has support over an area of order $\hbar$ in phase space. It evolves like the classical density to cover phase space exponentially fast. This occurs on a time scale that is roughly \cite{BB}
\EQ{
T\thicksim\lambda^{-1}\log\frac{I}\hbar\ ,
}
where $I$ is a characteristic macroscopic action scale. The important point here, is that $T$ only depends on $\hbar$ logarithmically and so the spreading of even a minimal uncertainty state is unexpectedly fast. A much discussed example is the chaotic rotational motion of Hyperion one of the moons of Saturn \cite{Berry,PZ,Zurek2}. It has a Lyapunov exponent of roughly $(100\,\text{days})^{-1}$ and $I/\hbar\sim10^{58}$ is roughly its angular momentum in units of $\hbar$. So, starting from a rotational coherent state, the rotational wave function would become macroscopically spread out after only about $40$ years! Needless-to-say, this kind of macroscopic quantum state cannot be the right description and there has to be some mechanism to explain why Newton's Laws work fine. This shows that wave function spreading in the quantization of a classically chaotic system cannot be ignored in any discussion of the quantum-to-classical transition. One must go beyond, or at least provide a different context for, Ehrenfest's Theorem in order to derive classical mechanics from quantum mechanics.

In searching for a detailed theory of the quantum-to-classical limit, there is a promising approach based on the idea that a system behaves as if it was being continuously measured by its environment. The resulting theory of {\it quantum trajectories\/} is very successful at describing the state of a microscopic system, like a single atom, conditioned on the results of continuous measurements made on the electromagnetic field with which it interacts.\footnote{The subject, pioneered by Carmichael \cite{C1}, has a large literature: see the introductory articles \cite{JacobsSteck,PK,W1} and book \cite{MW} and references therein.} Could the same theory be used to describe a macroscopic system interacting with its environment? Unfortunately  there is a ambiguity because the resulting {\it conditioned\/} dynamics depends on how, in the example of the atom, the electromagnetic field is measured rather than being determined intrinsically by the atom. For instance, if the arrival time of the photons are measured, then the conditioned state of the atom evolves according to a jump process while if the photons are measured using homodyne detection, then the atom evolves more smoothly and in a certain limit according to a diffusion process \cite{W1,MW}. So the conditioned dynamics does not describe the intrinsic dynamics of the atom, but rather the state of the atom conditioned by the state of the measuring device. This is a sophisticated example of {\it complementarity\/} in the context of a continuous measurement.

One could take the same idea to describe a macroscopic system and fix the ambiguity by 
hand. For example, a convenient choice gives rise to quantum trajectories that are described by a stochastic Schr\"odinger equation known as {\it quantum state diffusion\/}. This equation has remarkable properties: wave packets localize on microscopic scales and so Ehrenfest's Theorem now consistently applies and wave packets follow classical trajectories, even in a classically chaotic system \cite{Brun1,Brun2,Brun3,Rigo1,Gisin1,Gisin2,Bhattacharya:1999gx,BHJ1,GADBH,Habib:1998ai}. The stochastic element accounts for the noise that is negligible for macroscopic systems, but as they become less macroscopic gives rise to the familiar random walk of Brownian motion. This is very impressive indeed: in this scenario classical mechanics really does emerge from quantum mechanics! However, the issue of basis ambiguity is unresolved and begs the question: is there a more natural way to fix the basis and will the resulting dynamics share the good properties of quantum state diffusion? 

A key notion in our approach is the idea of the {\it frame of reference\/} associated to a subsystem of a large quantum system, what we call the {\it subsystem frame\/}. Different frames will necessarily associate different states to the same system. Third party frames, external to the system of interest, would describe the state using the Schr\"odinger equation whereas a subsystem frame would describe the state using a combination of the Schr\"odinger equation and the Born rule. So the randomness of quantum mechanics arises in a subsystem frame.\footnote{The idea that the quantum state depends on the {\it frame of reference\/} of an observer is an idea that have surfaced in several contexts in quantum mechanics; including super-selection rules \cite{AS,DBRS,BRS1,BRS2} but also in discussions of interpretations, e.g.~\cite{Rov,Bene:1997kk,Sud1,Hollowood:2013cbr,Hollowood:2013xfa,Bruk,Faist}.} A classical thought experiment that illustrates the idea of subsystem frames is ``Wigner's friend''.  The friend $F$ measures a qubit $c_1\ket{+}+c_2\ket{-}$. In Wigner's frame the total state is obtained by solving the Schr\"odinger equation: $\ket{\psi}=c_1\ket{+}\ket{F_+}+c_2\ket{-}\ket{F_-}$. On the other hand, the state in the friend's own subsystem frame is either $\ket{\psi_1}=\ket{+}\ket{F_+}$ or $\ket{\psi_2}=\ket{-}\ket{F_-}$ with probabilities $|c_1|^2$ and $|c_2|^2$, respectively. There is no contradiction because the states $\ket{\psi_i}$ are not orthogonal to $\ket{\psi}$. Positing frames of reference unifies the Copenhagen interpretation, which describes the state in the friend's frame, with the many worlds interpretation, which describes the state in Wigner's frame.

The frames that are important for describing the macroscopic world are necessarily those that correspond to spatially localized, macroscopic subsystems. The state of a subsystem can be given a ensemble interpretation which is the key to unlocking the stochastic dynamics but the basis ambiguity we have highlighted arises because there are many inequivalent ensembles. So an important part of the story is to explain why there is a special, preferred ensemble. This will lead us to a version of theory of continuous measurement that we call {\it Born unravelling\/}, first defined by Di\'osi \cite{Diosi1,Diosi2} but related to the Schmidt histories described by Paz and Zurek \cite{Paz:1993tg} and having a promising phenomenology for the quantum-to-classical transition that was investigated in \cite{Gisin2,BH1,BH2,SH}.\footnote{Histories, or trajectories, based on the Schmidt decomposition have been studied by various authors, including \cite{Albrecht:1992uc,Kent:1996kx,Sud1,Hollowood:2013cbr,Hollowood:2013xfa}.}

\section{Entanglement or randomness}

We begin by analysing a simple measurement. The key idea is that interactions between two subsystems creates entanglement but, from the view of one of the subsystem frames---the state within the state---entanglement is experienced as a random non-entangled state.

Suppose a device $M$ measures a qubit. The initial state of the combined system is the separable state
\EQ{
\ket{\psi(0)}=\ket{M_0}\big(c_1\ket{+}+c_2\ket{-}\big)\ .
\label{sss}
}
$M$ then interacts with the qubit and the state becomes entangled
\EQ{
\ket{\psi(t)}=c_1\ket{M_+(t)}\ket{+}+c_2\ket{M_-(t)}\ket{-}\ .
\label{scg}
}
As time evolves, the two states $\ket{M_{\pm}(t)}$, with $\ket{M_\pm(0)}=\ket{M_0}$, become orthogonal after a short time scale $\delta t$. The subsystem $M$ defines a frame in which for any observable acting only within the Hilbert space of $M$ is indistinguishable from an ensemble defined by the reduced density matrix:
\EQ{
\rho(\delta t)=|c_1|^2\ket{M_+}\bra{M_+}+|c_2|^2\ket{M_-}\bra{M_-}\ ,
}
where $\ket{M_\pm}\equiv\ket{M_\pm(\delta t)}$. In $M$'s frame the state becomes random, i.e.~is a member of the ensemble, $\ket{M_\pm}$ with probability $|c_1|^2$ or $|c_2|^2$, respectively, as dictated by Born's rule. The uniqueness of the ensemble requires that $|c_1|^2\neq|c_2|^2$ and the fact that $\ket{M_\pm}$ are orthogonal. Note that the members of the ensemble lift to the pure states of the total system $\ket{M_\pm}\ket{\pm}$, the components of the Schmidt decomposition of the final state \eqref{scg} at $\delta t$.

Of course this is not a realistic model of the measuring process in quantum mechanics, but it illustrates the concepts that will be important; namely, a subsystem like $M$ defines a frame in which entanglement is replaced by randomness. In the following, we must be careful to distinguish which state we are referring to. In keeping with the convention of measurement theory, we shall call the state in $M$'s frame, so $\ket{M_\pm}$, the {\it conditioned\/} state (conditioned on the outcome of the experiment). On the other hand, the state of $M$ that is relevant to a third party, the density matrix $\rho(t)$, is the {\it unconditioned} state. In the example of Wigner friend, the friend describes $M$ via the conditioned state, while Wigner describes $M$ by the unconditioned state, until he opens the door, interacts and becomes entangled with his friend and learns the result.
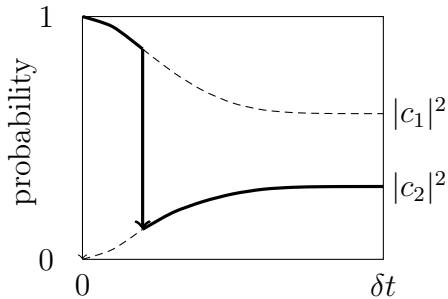
\begin{figure}[!h]
\begin{center}
\begin{tikzpicture}[yscale=0.17,xscale=0.4]
\draw[->] (0,0) -- (10,0) -- (10,19) -- (0,19) -- (0,0);
\node at (-1.2,19) (a1) {$1$};
\node at (-1.2,0) (a1) {$0$};
\node[rotate=90] at (-2,10) (a1) {probability};
\node at (10,-2) (d1) {$\delta t$}; 
\node at (0,-2) (a3) {$0$};
\node at (11.2,11.4005) {$|c_1|^2$};
\node at (11.2,5.6) {$|c_2|^2$};
\draw[very thick] plot[smooth] coordinates {(0, 19.)  (1, 18.2124)  (2, 16.4584)};
\draw[->,very thick]  (2, 16.4584) --  (2, 2.35062);
\draw[very thick] plot[smooth] coordinates { (2, 2.35062)  (3, 3.65956) (4,4.51829)  (5, 5.12001) (6, 5.48233)  (7, 5.6376)  (8, 5.68586)  (9,5.69741)  (10, 5.69961)};
\draw[densely dashed] plot[smooth] coordinates {(2, 16.4584)  (3, 14.6642)  (4, 13.2187)  (5,
   12.2299)  (6, 11.6976)  (7, 11.4838)  (8, 11.4189)  (9,11.4035)  (10, 11.4005)};
\draw[densely dashed] plot[smooth] coordinates {(0, 0)  (1, 0.773204)  (2, 2.35062) };
\end{tikzpicture}
\end{center}
\caption{\small A possible quantum trajectory for in the subsystem frame of $M$ where the state makes a jump between the eigenvectors of $\rho$. A stochastic average over the trajectories gives the probabilities $|c_i|^2$ at the end of the time interval.}
\label{f1}
\end{figure}

Another important point is that there is no macroscopic collapse of the wave function. As the state \eqref{scg} evolves the eigenvalues $p_a(t)$ of $\rho(t)$ vary smoothly: they start out as $p_1(0)=1$ and $p_2(0)=0$ and then they evolve continuously in time until $p_1(\delta t)=|c_1|^2$ and $p_2(\delta t)=|c_2|^2$, if $|c_1(t)|^2>|c_2(t)|^2$. Writing the eigenvectors as
\EQ{
\rho\ket{\phi_a(t)}=p_a(t)\ket{\phi_a(t)}\ ,
}
we have $\ket{\phi_1(0)}=\ket{M_0}$ and $\ket{\phi_1(\delta t)}=\ket{M_+}$ and $\ket{\phi_2(\delta t)}=\ket{M_-}$.
So the implication is that the state in subsystem frame of $M$ must evolve from $t=0$ as the eigenvector $\ket{\phi_1(t)}$ and then have the possibility to make a quantum jump discontinuously into the eigenvector $\ket{\phi_2(t)}$. So with the notion of a subsystem frame the existence of quantum jumps is inevitable: see figure \ref{f1}. The particular time evolution of the conditioned state is a {\it quantum trajectory\/}. A stochastic average of the conditioned state in the form $\ket{\psi}\bra{\psi}$ gives the unconditioned state, the density matrix $\rho$:
\EQ{
\mathscr E\big(\ket{\psi}\bra{\psi}\big)=\rho\ ,
}
where $\mathscr E(\cdot)$ denotes a stochastic average over trajectories. For present purposes, we do not need to specify the rules that govern the jumps in detail, all the matters are the rates averaged over the time interval and the final probabilities. However, we completeness we formulate a complete theory of the jumps in the next section. 

Before passing on, we remark that the jumps can only occur when the states $\ket{M_\pm(t)}$ are not orthogonal, in other words they cannot occur when the states become macroscopically distinct.

The simple model we have presented is not the answer to why a classical world results from quantum mechanics. The model is not robust: the states $\ket{M_\pm}$ are just assumed to be macroscopically distinct and only with exactly the right Hamiltonian will they turn out to to be the eigenvectors of the final density matrix of $M$. Any kind of perturbation, or measuring inefficiency will lead to the eigenvectors which are mixtures of the $\ket{M_\pm}$. However, the model points us in the right direction: it is the continual interactions of a macroscopic system $M$ with its environment that eventually will yield a robust derivation of the classical from the quantum.

\section{Born unravelling}

Now we take the lessons of the simple measurement and apply it more generally. $M$ is a macroscopic subsystem but instead of making specific measurements on a microscopic system, we now want to think about it interacting with its much larger environment.

\vspace{0.3cm}
\noindent
{\bf Quantum jumps.}~~~Again let us consider the two subsystems $M$ and $\cal E$, initially in a separable state 
\EQ{
\ket{\psi_0}=\ket{M_0}\ket{{\cal E}_0}\ .
}
The subsystems then interact for an interval of time $[0,\delta t]$ and the state becomes entangled
\EQ{
\ket{\psi(t)}=U_{t,0}\ket{\psi_0}=\sum_a\ket{M_a}\ket{{\cal E}_a}\ .
\label{wre}
}
The states $\ket{{\cal E}_a}$ and $\ket{M_a}$---implicitly time dependent---are orthogonal sets and so \eqref{wre} is the---generically unique---Schmidt decomposition of $\ket{\psi(t)}$. Note that for later convenience we have absorbed the weights into the definition of the state $\ket{M_a}$ (the $\ket{{\cal E}_a}$ are normalized) so that the conditioned state at time $t$ in the $M$ subsystem frame is one of the $\ket{M_a}$ with a probability equal to the norm
\EQ{
p_a=\bra{M_a}M_a\rangle\ .
\label{vbc}
}
For now, we make no assumption that the states $\ket{M_a}$ have any consistent classical interpretation: this will emerge much further into the discussion.

The probabilities $p_a$ are time dependent and so there must be jumps between the states in $M$'s subsystem frame (the conditioned state). The time derivative of the probability $p_a$ can be written,
\EQ{
\frac{dp_a}{dt}=\sum_b\big(\omega_{ab}-\omega_{ba}\big)\ ,\qquad \omega_{ab}=\bra{\psi}\Big(\frac{d\Pi_a}{dt}+\frac1{i\hbar}[\Pi_a,H]\Big)\Pi_b\ket{\psi}\ ,
}
where $\Pi_a$ are projectors onto the components of \eqref{wre}. Note that $\omega_{ab}-\omega_{ba}$ is real and probability is manifestly conserved $\sum_a\dot p_a=\sum_{ab}(\omega_{ab}-\omega_{ba})=0$. Using the orthogonality of the bases $\ket{M_a}$ and $\ket{{\cal E}_a}$ and taking the total Hamiltonian $H=H_M+M_{\cal E}+H_I$, we have
\EQ{
\omega_{ab}=\frac1{i\hbar}\bra{M_a}\bra{{\cal E}_a}H_I\ket{M_b}\ket{{\cal E}_b}\ ,
}
where $H_I$ is the interaction Hamiltonian between $M$ and ${\cal E}$. The expression for $\dot p_a$ above naturally suggests that there is a jump process with instantaneous transition rates:
\EQ{
r_{\ket{M_b}\to\ket{M_a}}=\frac1{p_b}[\omega_{ab}-\omega_{ba}]^+\ ,
\label{rte}
}
where $[x]^+=\text{max}(x,0)$.\footnote{Transition rates of this form were first written down in a different context by Bell \cite{Bell:2004suqm} in his theory of {\it beables\/} and then studied in the context of {\it modal interpretations\/};  for example see \cite{Sud1,BacciagaluppiDickson:1999dmi}.} Note the intuitive reason for the $1/p_b$ factor is because if $p_b\to0$ then the system must have a high rate of transition to leave the state $\ket{M_b}$. The rate above is not unique because one can add to $\omega_{ab}$ any $\mu_{ab}$ where $\sum_b(\mu_{ab}-\mu_{ba})=0$, however, the expression \eqref{rte} is the simplest choice that meshes with quantum mechanics (there is no obvious expression for $\mu_{ab}$) with transitions only going in one direction. In the end, the rest of our analysis is insensitive to this ambiguity because we only need the coarse-grained integrated jump rate $\ket{M_0}\to\ket{M_a}$ across the time step and this is simply $p_a(\delta t)$.

\vspace{0.3cm}
\noindent
{\bf Coarse graining and a quantum channel.}~~~The interaction time $\delta t$ is microscopically small and so it makes sense to coarse grain the jump process across the time interval and describe the dynamics as a discrete process (and then take a continuum limit eventually). The coarse graining is simple because the initial state is non-entangled; hence, $p_a(0)=\delta_{a1}$ with $\ket{M_1(0)}=\ket{M_0}$ and the jump process leads to the final states $\ket{M_a}\equiv\ket{M_a(\delta t)}$ with probabilities $p_a\equiv p_a(\delta t)=\bra{M_a}M_a\rangle$. These final states become individual decohered branches for the subsequent evolution.

We can describe the interaction across the interval $[0,\delta t]$ between the two systems from the third party frame---the unconditioned dynamics---in terms of the density matrix of $M$, as a transformation
\EQ{
\rho_0\longrightarrow \rho(\delta t)=\sum_aK_a\rho_0K_a^\dagger\ ,
\label{klo}
}
with $\rho_0=\ket{M_0}\bra{M_0}$ and 
where the operators 
\EQ{
K_a=\bra{{\cal E}_a}U_{\delta t,0}\ket{{\cal E}_0}\ ,
\label{rff}
}
are complete in the sense that
\EQ{
\sum_aK_a^\dagger K_a=1\ ,
\label{nm}
}
which follows from the unitarity of the underlying time evolution. 

Note that the conditioned state is one of the components in the sum \eqref{klo}, i.e.~a pure state:
\EQ{
K_a\rho_0K_a^\dagger\qquad\text{or}\qquad K_a\ket{M_0}\ ,
}
as a density matrix or state vector, respectively. These are the decoherent branches.

In the language of quantum information theory, the $K_a$ are Krauss operators that define a {\it quantum channel\/}, a map between density matrices:
\EQ{
\rho\longrightarrow\sum_aK_a\rho K_a^\dagger\ .
\label{dms}
}
The condition \eqref{nm} ensures that the channel is trace preserving. It is well known that the Krauss operators are not unique because \eqref{dms} is invariant under unitary $\text{U}(N)$ transformations $K_a\to\sum_bU_{ab}K_b$, $UU^\dagger=1$. However, the subsystem frame has picked out a specific basis, associated to the eigenbasis of $\rho(t)$, or Schmidt decomposition of the total state. This condition can be expressed as an orthogonality condition on the Krauss operators:
\EQ{
\Tr(K_a\rho_0K_b^\dagger)=\bra{M_0}K_b^\dagger K_a\ket{M_0}=p_a\delta_{ab}\ ,
\label{xll}
}
which for generic $p_a$ fixes the ${\text U}(N)$ symmetry.\footnote{Actually, there is still the abelian subgroup $\text{U}(1)^N$ unfixed but this rotates the Krauss operators $K_a$ by physically irrelevant phases.}
\pgfdeclareimage[interpolate=true,width=10cm]{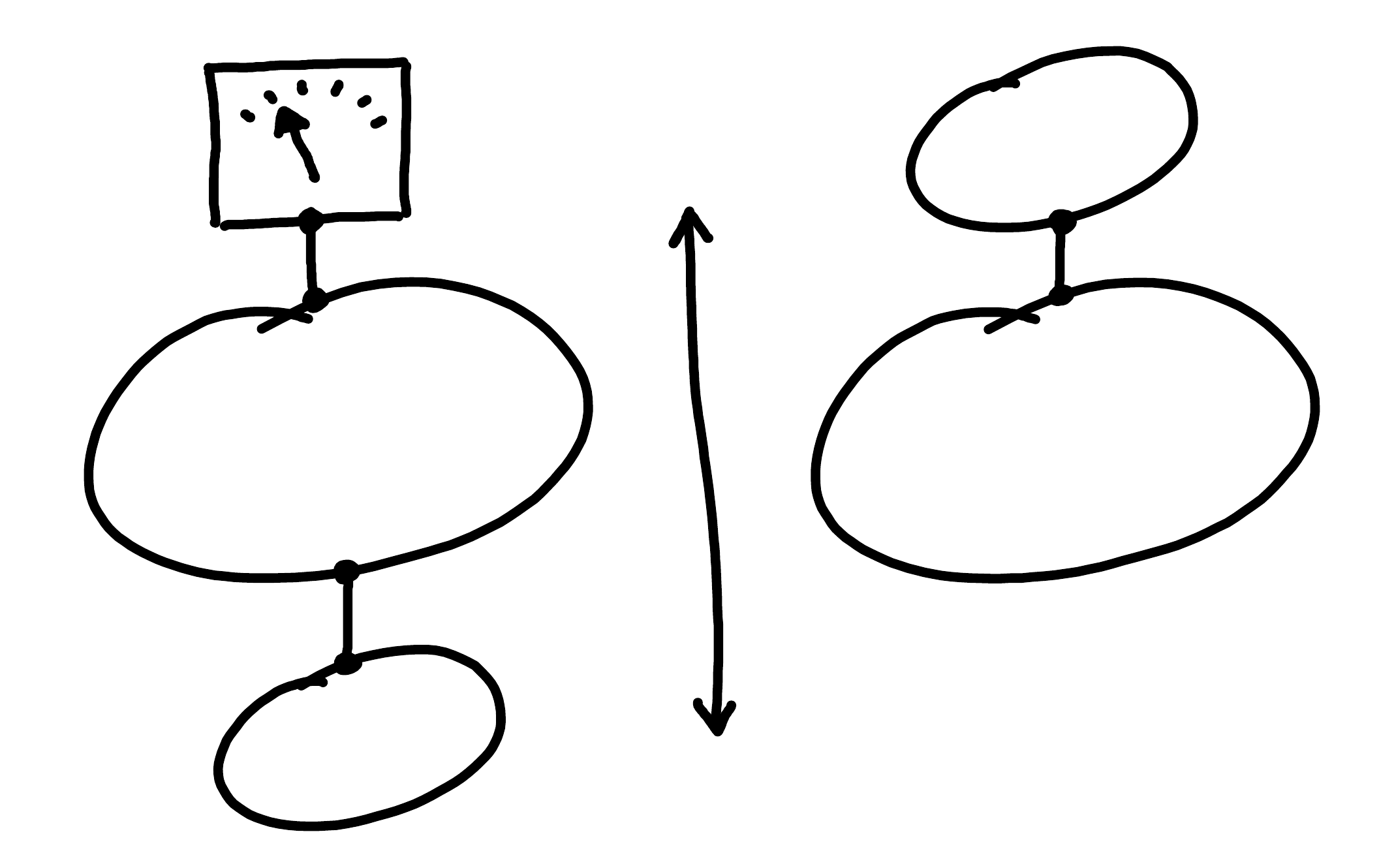}{fig2}
\begin{figure}[!h]
\begin{center}
\begin{tikzpicture}[scale=0.8]
\pgftext[at=\pgfpoint{0cm}{0cm},left,base]{\pgfuseimage{fig2}} 
\node at (7.6,5.3) {\Large$M$};
\node at (10.3,4.55) (a2) {\small branch-Schmidt};
\node at (10.7,4) {\small decomp.};
\draw[->] (a2) -- (7.9,4.55);
\node at (7.65,3.2) {\Large${\cal E}$};
\node at (2.4,3.2) {\Large${\cal E}$};
\node at (2.6,0.9) {\Large$M$};
\node at (5,5.1) {macro};
\node at (5.2,0.7) {micro};
\node at (2,-0.5) {(a) quantum trajectories set up};
\node at (8,-0.5) {(b) present set up};
\node at (0.4,4.3) (a1) {$\ket{{\cal E}_a}$};
\node at (-1.5,0) {\phantom{}};
\draw[->] (a1) -- (1.95,4.3);
\end{tikzpicture}
\caption{\footnotesize  (a) In the conventional continuous measurement/quantum trajectories formalism, $M$ is a microscopic system interacting with its environment $\cal E$ the latter being monitored by an external measuring device. The r\^ole of the measuring device is simply to fix a basis of environmental states ${\cal E}_a$---a measurement scheme---which then determines the basis of Krauss operators for the conditioned dynamics of $M$ with respect to the measuring device (an unravelling). The basis can be changed by changing what is measured. (b) In the present set up, $M$ is a macroscopic system and there is no external measuring device. The basis is fixed by continually applying Born's rule to the branch-Schmidt basis determined by the way $M$ interacts with ${\cal E}$.}
\label{f2} 
\end{center}
\end{figure}
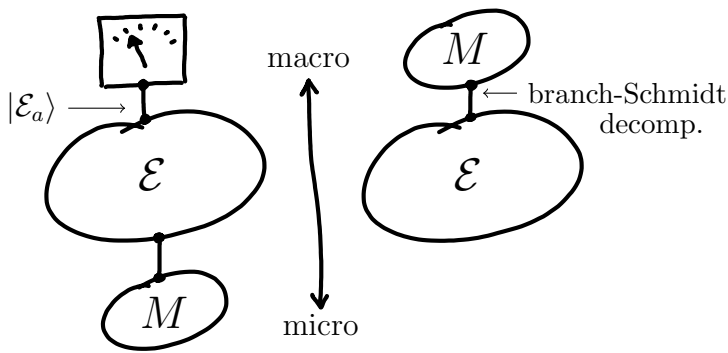

The fact that we can describe the time evolution of the conditioned state in terms of a quantum channel relates our approach to the subject of {\it continuous measurement\/} and  
{\it quantum trajectories\/} (see e.g.~\cite{C1,JacobsSteck,PK,W1,MW}). However, there is an important conceptual difference: see figure \ref{f2}. In the theory of quantum trajectories $M$ is a microscopic system and ${\cal E}$ is its environment. For example, $M$ could be an atom and ${\cal E}$ could be the electromagnetic field. In this case, the basis of Krauss operators is determined by how the environment ${\cal E}$ is then measured by an external measuring device. The measurement on ${\cal E}$ picks out a preferred set of orthonormal states of the environment $\ket{{\cal E}_a}$ which determines a preferred set of Krauss operators as in \eqref{rff}. So different ``measurement schemes'' give rise to a different basis of Krauss operators and  inequivalent dynamics of the conditioned state---what are known as {\it unravellings\/}. In the theory of quantum trajectories, the r\^ole of the external measuring device is just to fix the basis of the Krauss operators. On the contrary, in the present discussion $M$ is a macroscopic system and ${\cal E}$ is not being measured by an external system. The idea is that $M$ defines a frame of reference, and in this frame the state is conditioned intrinsically according what we will discover is a dynamical refinement of the Schmidt decomposition.

\vspace{0.3cm}
\noindent
{\bf Continual interactions.}~~~Now let us consider what happens when there are two systems ${\cal E}^1$ and ${\cal E}^2$ and $M$ now interacts with ${\cal E}^1$ for an interval of time $[0,\delta t]$ and never interacts with it again. Then $M$ interacts with the second system ${\cal E}^2$ over the next interval $[\delta t,2\delta t]$.

After the first measurement, $M$ and ${\cal E}^1$ are in the unconditioned state
\EQ{
\sum_{a_1}K_{a_1}\ket{M_0}\ket{{\cal E}^1_{a_1}}\ .
}
Since the system ${\cal E}^1$ disperses, i.e.~never interacts with $M$ (or other parts of the environment) again, the conditioned states $K_{a_1}\ket{M_0}$ relevant to the frame of reference $M$ picked out as the eigenvectors of the density matrix of $M$, remain decoherent for all future times. Consequently, they can be treated independently from the point-of-view of the second interaction with ${\cal E}^2$. For one of the components $K_{a_1}\ket{M_0}$, the result of the second interaction leads to the conditioned states that we write as
\EQ{
\ket{M_{a_2a_1}}=K_{a_2(a_1)}K_{a_1}\ket{M_0}\ .
}
realized with a probability given again by the norm
\EQ{
p_{a_2a_1}=\bra{M_{a_2a_1}}M_{a_2a_1}\rangle\ .
}
The $(a_1)$ dependence of the second Krauss operator indicates that the basis has been chosen so that the states are orthogonal with a given fixed value of $a_1$:
\EQ{
\bra{M_{a_2a_1}}M_{b_2a_1}\rangle=p_{a_2a_1}\delta_{a_2b_2}\ .
}
In particular, notice that there is no need for orthogonality between states with different values of $a_1$: these states are completely decoherent for $t>\delta t$.  What we mean by this, is that these states lift to states of the total system 
\EQ{
\ket{M_{a_2a_1}}\longrightarrow \ket{M_{a_2a_1}}\ket{{\cal E}^1_{a_1}}\ket{{\cal E}^2_{a_2}}\ ,
}
which are orthogonal for both labels $a_1$ and $a_2$, and two such states labelled by $a_2a_1$ and $b_2b_1$ for $a_1\neq b_1$ are not coupled by the interaction Hamiltonian between $M$ and ${\cal E}$ because for $t\in[\delta t,2\delta t]$
\EQ{
\bra{{\cal E}^1_{a_1}}H_I\ket{{\cal E}^1_{b_1}}=0 \qquad \text{for}\qquad a_1\neq b_1\ .
}
As far as $M$ is concerned the trajectories labelled by different labels $a_1$ can never interfere with each other in any experiment local to $M$ and the jump rates \eqref{rte}
\EQ{
r_{\ket{M_{a_2a_1}}\to\ket{M_{b_2b_1}}}=0\qquad\text{for}\qquad a_1\neq b_1\ .
}

For fixed $a_1$, the $K_{a_2(a_1)}$ are a set of bona-fide Krauss operators:
\EQ{
\sum_{a_2}K_{a_2(a_1)}^\dagger K_{a_2(a_1)}=1\ .
}
The unconditioned state of $M$ after both interactions can therefore be written
\EQ{
\rho(2\delta t)=\sum_{a_1a_2}K_{a_2(a_1)}K_{a_1}\rho_0K_{a_1}^\dagger K_{a_2(a_1)}^\dagger\ ,
}
which, itself, is a quantum channel with Krauss operators $K_{a_2(a_1)}K_{a_1}$:
\EQ{
\sum_{a_1a_2}\big(K_{a_2(a_1)}K_{a_1}\big)^\dagger K_{a_2(a_1)}K_{a_1}=\sum_{a_1}K_{a_1}^\dagger\Big(\sum_{a_2}K_{a_2(a_1)}^\dagger K_{a_2(a_1)}\Big) K_{a_1}=\sum_{a_1}K_{a_1}^\dagger K_{a_1}=1\ .
}

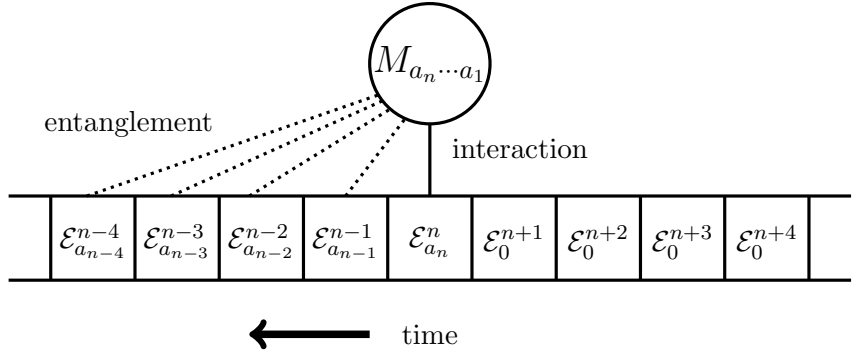
\begin{figure}[!h]
\begin{center}
\begin{tikzpicture}[scale=0.8]
\draw[very thick] (7,5) circle (1cm); 
\node at (7,5) {\large $M_{a_n\cdots a_1}$};
\begin{scope}[scale=1.4]
\draw[very thick] (0,1) -- (10,1);
\draw[very thick] (0,2) -- (10,2);
\draw[very thick] (0.5,1) -- (0.5,2);
\draw[very thick] (1.5,1) -- (1.5,2);
\draw[very thick] (2.5,1) -- (2.5,2);
\draw[very thick] (3.5,1) -- (3.5,2);
\draw[very thick] (4.5,1) -- (4.5,2);
\draw[very thick] (5.5,1) -- (5.5,2);
\draw[very thick] (6.5,1) -- (6.5,2);
\draw[very thick] (7.5,1) -- (7.5,2);
\draw[very thick] (8.5,1) -- (8.5,2);
\draw[very thick] (9.5,1) -- (9.5,2);
\node at (1,1.45) {\small ${\cal E}^{n-4}_{a_{n-4}}$};
\node at (2,1.45) {\small ${\cal E}^{n-3}_{a_{n-3}}$};
\node at (3,1.45) {\small ${\cal E}^{n-2}_{a_{n-2}}$};
\node at (4,1.45) {\small ${\cal E}^{n-1}_{a_{n-1}}$};
\node at (5,1.45) {\small ${\cal E}^n_{a_{n}}$};
\node at (6,1.45) {\small ${\cal E}^{n+1}_{0}$};
\node at (7,1.45) {\small ${\cal E}^{n+2}_{0}$};
\node at (8,1.45) {\small ${\cal E}^{n+3}_{0}$};
\node at (9,1.45) {\small ${\cal E}^{n+4}_{0}$};
\end{scope}
\draw[very thick] (7,2.8) -- (7,4);
\draw[very thick,dotted] (1.3,2.8) -- (6.2,4.5);
\draw[very thick,dotted] (2.7,2.8) -- (6.25,4.4);
\draw[very thick,dotted] (4,2.8) -- (6.38,4.27);
\draw[very thick,dotted] (5.6,2.8) -- (6.6,4.1);
\draw[line width=1mm,->] (6,0.5) -- (4,0.5); 
\node at (7,0.5) {\small time};
\node at (8.5,3.6) {\small interaction};
\node at (2,4) {\small{entanglement}};
\end{tikzpicture}
\caption{\footnotesize  The ticker-tape paradigm for the environment that lies behind the Born-Markov approximation. In each time interval $\delta t$, the system $M$ interacts with a fresh bit of the environment and becomes entangled with it. These parts of the environment then disperse to leave only their entanglement and no further interaction. This continually decoheres the states of $M$. The state of $M$ becomes conditioned on the term picked out in the Schmidt decomposition with each part of the environment in accordance with Born's rule. Note that the picture of the tape is not literal and the subsystems ${\cal E}^n$ are to be viewed as ``logical'' subsystems rather than spatially localized subsystems.}
\label{f3} 
\end{center}
\end{figure}

Clearly, this whole set up can be generalized to the case of many systems ${\cal E}^n$ which interact with $M$ in non-overlapping time intervals $[(n-1)\delta t,n\delta t]$. In this picture, the environment acts as a kind of ticker-tape, where in each time interval the system is presented with a fresh bit of the environment and previous correlations are dispersed  to the extent that there is no back reaction on $M$: see figure \ref{f3}. Note that the subsystems ${\cal E}^n$ are not necessarily identified as localized subsystems of ${\cal E}$, in general we can expect the correlations with $M$ to become widely distributed in ${\cal E}$ and so the subsystems ${\cal E}^n$ are ``logical'' subsystems.

As long as a given component disperses and never interacts with $M$ or the rest of ${\cal E}$ again, then the unconditioned state can be expressed as the convolution of many quantum channels:
\EQ{
\rho(n\delta t)=\sum_{a_1\cdots a_n}K_{a_n(a_{n-1}\cdots a_1)}\cdots K_{a_2(a_1)}K_{a_1}\rho_0K_{a_1}^\dagger K_{a_2(a_1)}^\dagger \cdots K_{a_n(a_{n-1}\cdots a_1)}^\dagger 
\ .
}
In the subsystem frame of $M$, the state is conditioned, equal to one of the components
\EQ{
\ket{M_{a_n\cdots a_1}}=K_{a_n(a_{n-1}\cdots a_1)}\cdots K_{a_2(a_1)}K_{a_1}\ket{M_0}\ .
}
At each level, the freedom to rotate the basis of Krauss operators is fixed by the orthogonality condition
\EQ{
\bra{M_{a_na_{n-1}\cdots a_1}}M_{b_na_{n-1}\cdots a_1}\rangle=p_{a_n\cdots a_1}\delta_{a_nb_n}\ .
\label{prb}
}
where $p_{a_n\cdots a_1}$ is the probability of the state.

We have described how the conditioned and non-conditioned states of a system $M$ behave when it interacts with a series of other subsystems in the approximation that each component ${\cal E}^n$ has its own time interval interacting with $M$ after which it disperses never to interact again. If we think of the union of ${\cal E}^n$ as being the environment ${\cal E}$ of $M$, then entanglements between $M$ and ${\cal E}$ established during the interactions are carried away and dispersed widely in the environment. This is the crux of the Born-Markov approximation and, as we have seen, leads to an effective unconditioned dynamics for $M$ that can be described as a series of quantum channels acting on $M$'s density matrix.

The Born unravelling we have described at this discrete level was first defined by Paz and Zurek \cite{Paz:1993tg}. In that reference, the idea was to make projections onto the Schmidt basis at time intervals greater than the decoherence time, here $\delta t$. For us the interpretation is different, there is no ad-hoc projection of the total state at times greater than the decoherence time, rather the conditioned state evolves in a piecewise continuous way, i.e.~with jumps. The time intervals for us are just a convenience rather than a necessity. It is interesting Paz and Zurek showed that if the conditioned states of $M$ are lifted to the total system by using the Schmidt decomposition, $\ket{M_{a_n\cdots a_1}}\to\ket{M_{a_n\cdots a_1}}\ket{{\cal E}^1_{a_1}}\cdots\ket{{\cal E}^n_{a_n}}$, the stochastic trajectories form a set of {\it consistent histories\/} \cite{Griffiths,GellMann:1992kh,Omnes:1992ag}. This latter property means that probabilities can consistently be assigned to the histories something that has been manifest in the quantum trajectories formalism.  

\vspace{0.3cm}
\noindent
{\bf Branch-Schmidt decomposition.}~~~The Born unravelling involves a particular kind of decomposition of the total state of the system that we want to highlight because it is {\it not\/} the Schmidt decomposition of the total state. If we take time $t=n\delta t$, the total state is
\EQ{
\ket{\Psi(n\delta t)}=\sum_{a_n\ldots a_1}\ket{M_{a_n\cdots a_1}}\ket{{\cal E}^1_{a_1}}\cdots\ket{{\cal E}^n_{a_n}}\ket{{\cal E}^{n+1}_0}\cdots\ .
\label{gxx}
}
It is not the Schmidt decomposition for the tensor product $\BH_M\otimes\BH_{\cal E}$ because the states $\ket{M_{a_n\cdots a_1}}$ are not orthogonal with respect to the string label $a_n\cdots a_1$, only for the last label $a_n$. However it does consist of a sum over the Schmidt decompositions of each {\it branch\/} labelled by $a_{n-1}\cdots a_1$.

In fact, the decomposition in \eqref{gxx} is naturally associated to a dynamical refinement of the Schmidt decomposition that takes into account the decoherence of $M$ induced by its interaction with ${\cal E}$. The idea is that the interaction Hamiltonian between $M$ and ${\cal E}$ defines effective super-selection sectors in $M$'s subsystem frame. Let us denote the interaction Hamiltonian $H_I$ and define the instantaneous orthogonal decomposition of the environmental Hilbert space
\EQ{
\BH_{\cal E}=\bigoplus_u\BH^{(u)}_{\cal E}\ ,
}
which is block diagonal with respect to $H_I$ meaning that for $u\neq v$
\EQ{
\bra{\phi}H_I\ket{\chi}=0\ ,\quad \forall\quad\ket{\phi}\in\BH^{(u)}_{\cal E}\ ,\quad\ket{\chi}\in\BH^{(v)}_{\cal E}\ ,
\label{yww}
}
as operators acting in $\BH_M$.\footnote{More precisely, with $A=\bra{\phi}H_I\ket{\chi}$ the condition can be defined as $\sqrt{\Tr(A^\dagger A)}/\hbar<T^{-1}$ where $T$ is IR temporal cut off, larger than any other relevant time scale. This ensures that the probability of a transition is vanishingly small across a time step.} The significance of the \eqref{yww} is that it ensures that the transition rates \eqref{rte} vanish between the effective super-selection sectors. So decoherence breaks the ergodicity of the jump process we defined earlier.\footnote{This refinement of the Schmidt decomposition solves in a natural way a problem of using the pure Schmidt decomposition as conditioned states. The problem is that the eigenstates of $\rho$ can mix wildly when two eigenvalues are accidentally nearly degenerate, even when the states are macroscopically distinct \cite{Hollowood:2013xfa}. The solution here is that the mixing cannot happen for the branch-Schmidt decomposition because macroscopically distinct states lie in orthogonal super-selection sectors, are completely decoherent and can never mix.}

With respect to the decomposition, the total state $\ket{\Psi}$ can be split up into {\it decoherent branches\/} $\ket{\Psi^{(u)}}$ each of which can be written in terms of a Schmidt decomposition
\EQ{
\ket{\Psi}=\sum_u\ket{\Psi^{(u)}}\ ,\qquad \ket{\Psi^{(u)}}=\sum_i c_i^{(u)}\ket{\psi_i^{(u)}}\ket{\phi_i^{(u)}}\ .
}
Here the states $\ket{\phi_i^{(u)}}\in\BH^{(u)}_{\cal E}$ are orthonormal but the states $\ket{\psi_i^{(u)}}$ are only orthonormal in a given branch (i.e~given $u$):
\EQ{
\bra{\phi_i^{(u)}}\phi_j^{(v)}\rangle=\delta_{ij}\delta_{uv}\ ,\qquad\bra{\psi_i^{(u)}}\psi_j^{(u)}\rangle=\delta_{ij}\ .
}
In particular, there is no requirement that the states $\ket{\psi_i^{(u)}}$ and $\ket{\psi_j^{(v)}}$ in different branches are orthogonal. This manifests the intuition that the states in different branches $\ket{\psi_i^{(u)}}$ and $\ket{\psi_j^{(v)}}$ of $M$, for $u\neq v$, are decohered by the environment and do not need to be orthogonal. Note that the new decomposition, unlike the Schmidt decomposition, is not symmetrical between the two subsystems.

The state \eqref{gxx} at time $t=n\delta t$ has precisely this decomposition with branch label $u=a_{n-1}\cdots a_1$ and $i=a_n$. The condition \eqref{yww} is satisfied because the components of the environment ${\cal E}^j$, $j=1,\ldots,n-1$, cease to interact with $M$ for $t>(n-1)\delta t$.

\section{The continuum limit}

For realistic systems, the time scale $\delta t$ for interactions with different components of the environment is very fast on macroscopic scales and so on these scales it is valid to coarse grain and view the time intervals as essentially---but not strictly---infinitesimal $dt$. This allows one to formulate the dynamics in terms of (stochastic) differential equations rather than discrete quantum channels. 

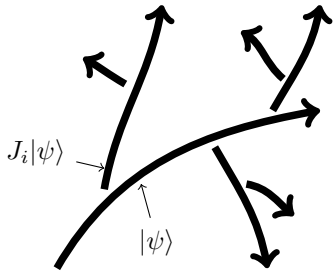
\begin{figure}[!h]
\begin{center}
\begin{tikzpicture}[scale=0.7]
\draw[line width=1mm,->] (0,0) to[out=60,in=-170] (5,3);
\draw[line width=1mm,->] (0.9,1.5) to[out=80,in=-100] (2,5);
\draw[line width=1mm,->] (3,2.3) to[out=-60,in=100] (4,0);
\draw[line width=1mm,->] (4.1,3) to[out=60,in=-100] (5,5);
\draw[line width=1mm,->] (1.3,3.5) to[out=150,in=-30] (0.5,4);
\draw[line width=1mm,->] (3.6,1.6) to[out=-10,in=140] (4.5,1);
\draw[line width=1mm,->] (4.3,3.6) to[out=140,in=-30] (3.5,4.5);
\node at (-0.4,2.2) (a1) {\footnotesize $J_i\ket{\psi}$};
\node at (1.9,0.5) (a2) {\footnotesize $\ket{\psi}$};
\draw[->] (a1) -- (0.9,1.9);
\draw[->] (a2) -- (1.6,1.6);
\end{tikzpicture}
\caption{\footnotesize  The conditioned state has a stochastic dynamics where jumps can occur into branches created by the operators $J_i$ with instantaneous rate $r_i=\bra{\psi}J_i^\dagger J_i\ket{\psi}$. The unconditioned state has a many worlds branch-like structure.}
\label{f10} 
\end{center}
\end{figure}

Let $\ket{\psi}$ be the instantaneous conditioned state. Over the next time interval $[t,t+dt]$, one must evolve the pure state density matrix $\rho_\psi=\ket{\psi}\bra{\psi}$ and then calculate its eigenstates.  Let us suppose there are $N+1$ eigenstates. One of them will be infinitesimally close to $\ket{\psi}$ and let us write the $N$ other states as
\EQ{
J_i\ket{\psi}\ ,
\label{laa}
} 
which defines the {\it branch creation operators\/} $J_i$. The picture here is that for each time step there is a certain probability for the conditioned state to jump into the branch created by $J_i$: see figure \ref{f10}. The subtle point is that they depend implicitly on the state $\ket{\psi}$ because the states \eqref{laa} have to be mutually orthogonal and orthogonal to $\ket{\psi}$, so
\EQ{
\bra{\psi}J_i\ket{\psi}=0\ ,\qquad\bra{\psi}J_i^\dagger J_j\ket{\psi}=r_i\delta_{ij}\ .
\label{ckk}
}
The $r_i$ are the rates for creating the $i^\text{th}$ branch $\ket{\psi}\to J_i\ket{\psi}$. The $N+1$ Krauss operators take the form
\EQ{
K_i=J_i\,\sqrt{dt}\qquad(i=1,2,\ldots,N)\ ,\qquad K_{N+1}=1+\frac1{i\hbar}H_\text{eff}\,dt\ .
}
The fact that the $K_i$ are order $\sqrt{dt}$ is needed because then $\bra{\psi}K_i^\dagger K_i\ket{\psi}=r_i\,dt$ is the correct probability for creating a branch during the time interval. The effective Hamiltonian $H_\text{eff}$---to be identified below---will include the actual Hamiltonian but have other terms and will not be Hermitian; indeed, the completeness relation \eqref{nm} specifies the non-Hermitian component of $H_\text{eff}$:
\EQ{
\sum_{i=1}^NJ_i^\dagger J_i+\frac1{i\hbar}\big(H_\text{eff}-H_\text{eff}^\dagger\big)=0\ .
}

The evolution of the unconditioned state $\rho$ of $M$ is then
\EQ{
\frac{\partial\rho}{\partial t}=\sum_{i=1}^{N+1}K_i\rho K_i^\dagger=\frac1{i\hbar}\big(H_\text{eff}\rho-\rho H_\text{eff}^\dagger\big)+\sum_{i=1}^NJ_i\rho J_i^\dagger\ .
\label{xkk}
}
One might be a bit puzzled at this point: the branch creation operators depend implicitly and non-linearly on the conditioned state $\ket{\psi}$, but the unconditioned evolution \eqref{xkk} should just depend on unconditioned state $\rho$. The resolution of this point will become apparent below where we show that when $H_\text{eff}$ is correctly identified \eqref{xkk} is actually independent of $\ket{\psi}$.
In order to write down the evolution of the conditioned state, it is useful to introduce the stochastic increments of a multi-component Poisson process $dN_i$, $i=1,2\ldots,N$, which are exclusive and are equal to 0 or 1.\footnote{But are infinitesimal because they only take the value 1 for an infinitesimal amount of time.} They satisfy the stochastic calculus rules
\EQ{
dN_i\,dN_j=\delta_{ij}\,dN_i\ ,\qquad dt\,dN_i=0\ , 
}
and have a stochastic average that involves the rate
\EQ{
\mathscr E(dN_i)=r_i\, dt\ ,\qquad \mathscr E\big(dN_i\,dN_j\big)=r_i\delta_{ij}\,dt\ .
}
The evolution of the conditioned state $\ket{\psi}$ of $M$ can then be written as the stochastic differential equation
\EQ{
d\ket{\psi}=\Big(\frac1{i\hbar}H_\text{eff}+\frac12\sum_{i=1}^Nr_i\Big)\ket{\psi}\,dt+\sum_{i=1}^N\Big(\frac{J_i}{\sqrt{r_i}}-1\Big)\ket{\psi}\,dN_i\ .
\label{hee}
}
This form preserves the normalization of $\ket{\psi}$. Note that the unconditioned evolution \eqref{xkk} can also be written as a stochastic average of the unconditioned evolution:
\EQ{
\rho+d\rho=\mathscr E\big((\ket{\psi}+d\ket{\psi})(\bra{\psi}+d\bra{\psi})\big)\ ,
}
and note that one must keep the $d\ket{\psi}\,d\bra{\psi}$ term because $dN_i\,dN_i=dN_i$ in stochastic calculus.

In order to find $H_\text{eff}$ and the $J_i$, we can match the unconditioned evolution \eqref{xkk} with the linear master equation of a subsystem interacting with an environment in the Born-Markov approximation which is usually written in manifestly trace preserving form as \cite{K1,GKS,L1}
\EQ{
\frac{\partial\rho}{\partial t}\equiv{\cal L}(\rho)=\frac1{i\hbar}[H,\rho]+\sum_{i=1}^N\big(A_i\rho A_i^\dagger-\frac12A_i^\dagger A_i\rho-\frac12\rho A_i^\dagger A_i\big)\ ,
\label{mms}
}
for a set of {\it Lindblad operators\/} $A_i$, $i=1,2,\ldots,N$. Cleary the first term corresponds to unitary dynamics which preserves the purity of the state. 

The question is how the standard Lindblad form in \eqref{mms} relates to \eqref{xkk}.
The form of the master equation \eqref{mms} is not unique because of two kinds of transformations on the operators $A_i$ that leave the form \eqref{mms} invariant. Firstly, shifting the Lindbald operators by constants, can be absorbed into a shift of the Hamiltonian:
\EQ{
A_i\to A_i+\lambda_i\ ,\qquad H\to H+\frac{i\hbar}2\sum_{i=1}^N(\lambda_i^*A_i-\lambda_iA_i^*)\ ,
\label{ssy}
}
for complex parameters $\lambda_i$. Secondly, there is an $\text{U}(N)$ symmetry of the Lindblad operators corresponding to $A_i\to\sum_j U_{ij}A_j$. So we can use the shift and unitary symmetry to relate the $A_i$ to the branch creation operators by imposing the conditions \eqref{ckk},\footnote{Here, and in the following $\VEV{\cdots}_\psi\equiv\bra{\psi}\cdots\ket{\psi}$.}
\EQ{
A_i=\VEV{A_i}_\psi+(U^{-1}\cdot J)_i\ .
\label{iss}
}
This fixes the unitary transformation $U$ up to the physically irrelevant abelian subgroup that rotates each $J_i$ by a phase. Note that $U$ depends implicitly on the state $\ket{\psi}$ and so the $J_i$ depend on the state via $\VEV{A_i}_\psi$ and $U$. The equality of \eqref{mms} and \eqref{xkk} then determines the effective Hamiltonian
\EQ{
H_\text{eff}=H+\frac{i\hbar}2\sum_{i=1}^N\Big\{\VEV{(U\cdot A)_i^\dagger}_\psi J_i-\VEV{(U\cdot A)_i}_\psi J_i^\dagger-J_i^\dagger J_i\Big\}\ .
\label{hef}
}
Notice that the last term is non-Hermitian.

Before continuing, it is worth pointing out the shift symmetry and $\text{U}(N)$ symmetry described above actually combine into the $\text{U}(N+1)$ symmetry that acts on the Krauss operators $K_i$, $i=1,2,\ldots,N+1$, in the canonical way. The $\text{U}(N)$ and shift symmetry are embedded in the $\text{U}(N+1)$ group as follows: in $([N]+[1])\times([N]+[1])$ block form as
\EQ{
\begin{pmatrix} U & 0 \\ 0 & 1 \end{pmatrix}\qquad\text{and}\qquad \begin{pmatrix} 1-\frac12\sum_{i=1}^N|\lambda_i|^2\,dt & \lambda_i\sqrt{dt} \\ -\lambda_i^*\sqrt{dt} & 1-\frac12\sum_{i=1}^N|\lambda_i|^2\,dt \end{pmatrix}\ ,
}
respectively. Note that the latter corresponds to $J_i\to J_i+\lambda_i$.

Now we can turn the philosophy around and derive the conditioned dynamics for any given master equation. This is the process of {\it unravelling\/} a master equation.
The continuum limit of the Born unravelling is actually a kind of {\it orthogonal unravelling\/} that was first formulated by Di\'osi \cite{Diosi1,Diosi2}. It is interesting that, in the conventional quantum trajectories formalism, the Born unravelling corresponds to the optimal measurement scheme in the sense of requiring the least information to keep track of the system \cite{Breslin1}. The intuition is that, because the jumps are always into orthogonal states, it requires the least number of quantum jumps of any measurement scheme. It is rather satisfying that the detailed microscopic model has led to such a distinguished form of unravelling.

For later use, the evolution of the expectation value in the conditioned state, takes the form
\EQ{
d\VEV{{\cal O}}_\psi=\Tr\big({\cal O}{\cal L}(\ket{\psi}\bra{\psi})\big)\,dt+\sum_{i=1}^N\VEV{J_i^\dagger({\cal O}-\VEV{{\cal O}}_\psi)J_i}_\psi\Big(\frac{dN_i}{r_i}-dt\Big)\ .
\label{uru}
}
Notice that if we perform a stochastic average that the second term vanishes and what results is the expectation in the unconditioned state $d\Tr(\rho{\cal O})=\Tr\big({\cal O}{\cal L}(\rho)\big)\,dt$.

It is interesting to compare Born with the Quantum State Diffusion (QSD) unravelling because the latter has proved popular in discussions of the quantum-to-classical transition \cite{Brun1,Brun2,Brun3,Rigo1,Gisin1,Gisin2,Bhattacharya:1999gx,BHJ1,GADBH,Habib:1998ai}. The conditioned evolution of the state for QSD takes a very similar form to the Born unravelling \eqref{hee}; indeed the effective Hamiltonian is identical:
\EQ{
\text{(QSD):}\qquad d\ket{\psi}=\frac1{i\hbar}H_\text{eff}\ket{\psi}\,dt+\sum_{i=1}^N J_i\ket{\psi}\,dW_i\ .
}
Here, $dW_i$ are the stochastic differentials of a set of Wiener processes that satisfy the Ito calculus $dW_i\,dW_j=\delta_{ij}\,dt$. The fact that the deterministic part of the dynamics of QSD and the Born unravelling are the same (up to a normalization factor) is important because it means that the resulting phenomenology of both unravellings will essentially be the same on macroscopic scales where the stochastic component is negligible. For QSD the evolution of an expectation value takes the form
\EQ{
\text{(QSD):}\qquad d\VEV{{\cal O}}_\psi=\Tr\big({\cal O}{\cal L}(\ket{\psi}\bra{\psi})\big)\,dt+\sum_{i=1}^N\VEV{J_i^\dagger{\cal O}+{\cal O}J_i}_\psi\,dW_i\ .
\label{uru2}
}

\section{Classical emergence}

Can the Born unravelling lead to classical mechanics in realistic situations? We have already remarked that the phenomenology of the Born unravelling should be very similar to QSD for macroscopic systems and the later emerges as the coarse grained description of the former when the jump rates $r_i$ are sufficiently large. This is already very encouraging because there is now ample evidence from both theoretical and numerical studies that classical mechanics does emerge from QSD \cite{Brun1,Brun2,Brun3,Rigo1,Gisin1,Gisin2,Bhattacharya:1999gx,BHJ1,GADBH,Habib:1998ai}. What is particularly impressive about this work is how all the phenomenology of chaotic classical system emerges in detailed numerical simulations. 

In order to focus the discussion, we will often use the simplest non-trivial model which is a single particle moving in one dimension interacting with a thermal environment: this is the well-studied model of quantum Brownian motion for which we refer to the textbook by Breuer and Petruccione \cite{Breuer1}. The effect of a typical thermal environment can be described in terms of a single Lindblad operator\footnote{Note that this involves an approximation that it strictly only valid at large enough temperature. The more fundamental master equation is not in Lindblad form and is non-Markovian.} 
\EQ{
A=\sqrt{\frac{4\gamma m kT}{\hbar^2}}x+i\sqrt{\frac{\gamma}{4mkT}}p\ .
\label{mnm}
}
In this case, there being only one Lindblad operator means the complications of the unitary transformation \eqref{iss} do not arise. In this expression, $T$ is the temperature of the environment and $\gamma$ is its relaxation rate. The master equation (for the unconditioned state) takes the form
\EQ{
\frac{\partial\rho}{\partial t}=\frac1{i\hbar}\big[\frac{p^2}{2m}+V(x)+\frac\gamma2(xp+px),\rho\big]+A\rho A^\dagger-\frac12A^\dagger A\rho-\frac12\rho A^\dagger A\ .
\label{mst}
}
For a macroscopic system one can ignore the ${\cal O}(p^2)$ terms from the Lindbald operators---at the expense of strict positivity---in which case this is the Cladeira-Leggett master equation \cite{CL1}. Note that the interaction with the environment also includes an additive shift in the usual Hamiltonian. The expectation values in the unconditioned state $\bar x_\rho=\Tr(x\rho)$ and $\bar p_\rho=\Tr(p\rho)$ satisfy the simple equations
\EQ{
\frac{d\bar x_\rho}{dt}=\frac{\bar p_\rho}m\ ,\qquad \frac{d\bar p_\rho}{dt}=-\Tr\big[\rho V'(x)\big]-2\gamma\bar p_\rho\ ,
\label{rvv}
}
but Ehrenfest's Theorem cannot be invoked since the unconditioned state is not localized.

\pgfdeclareimage[interpolate=true,width=8cm]{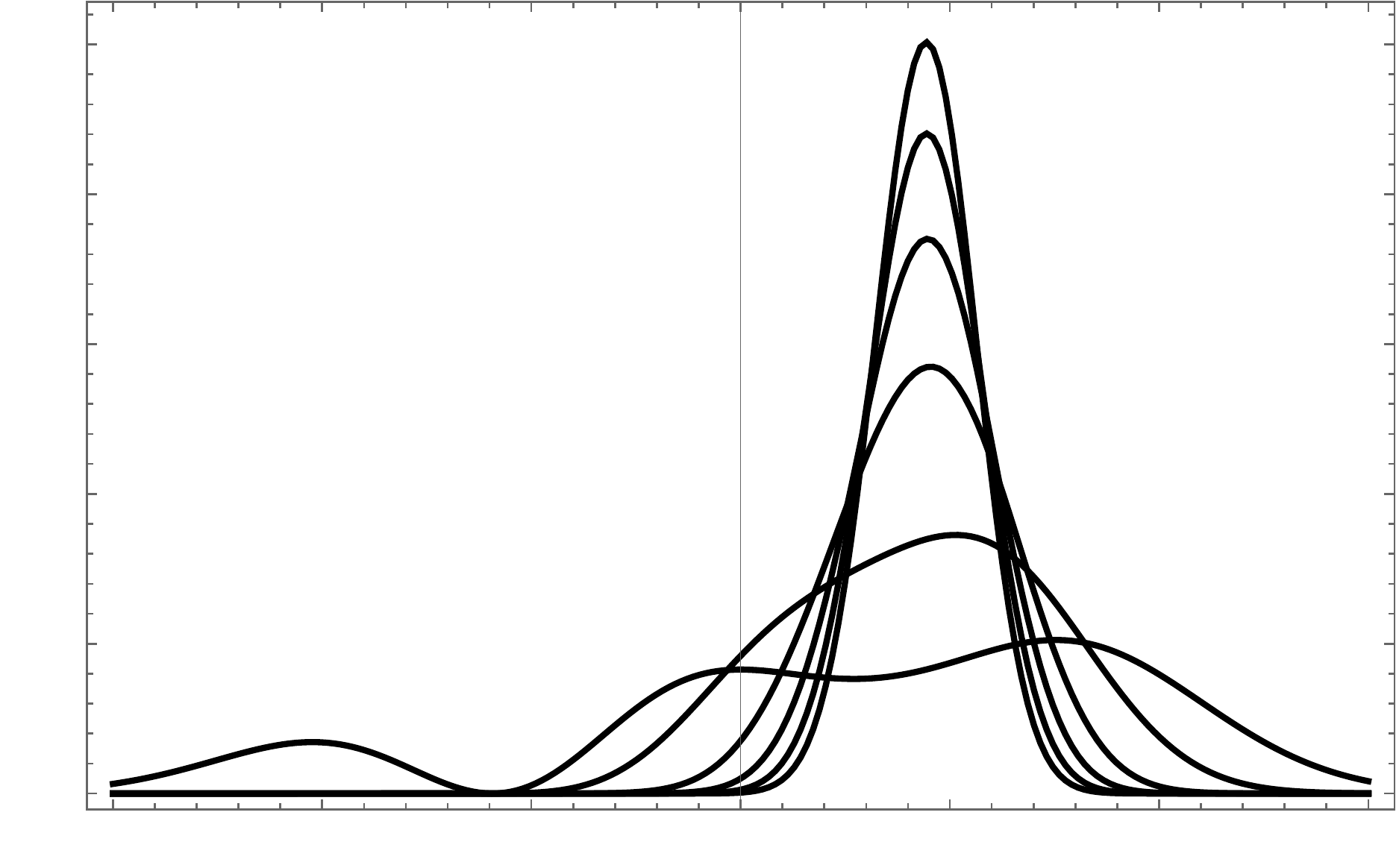}{fig4}
\pgfdeclareimage[interpolate=true,width=8cm]{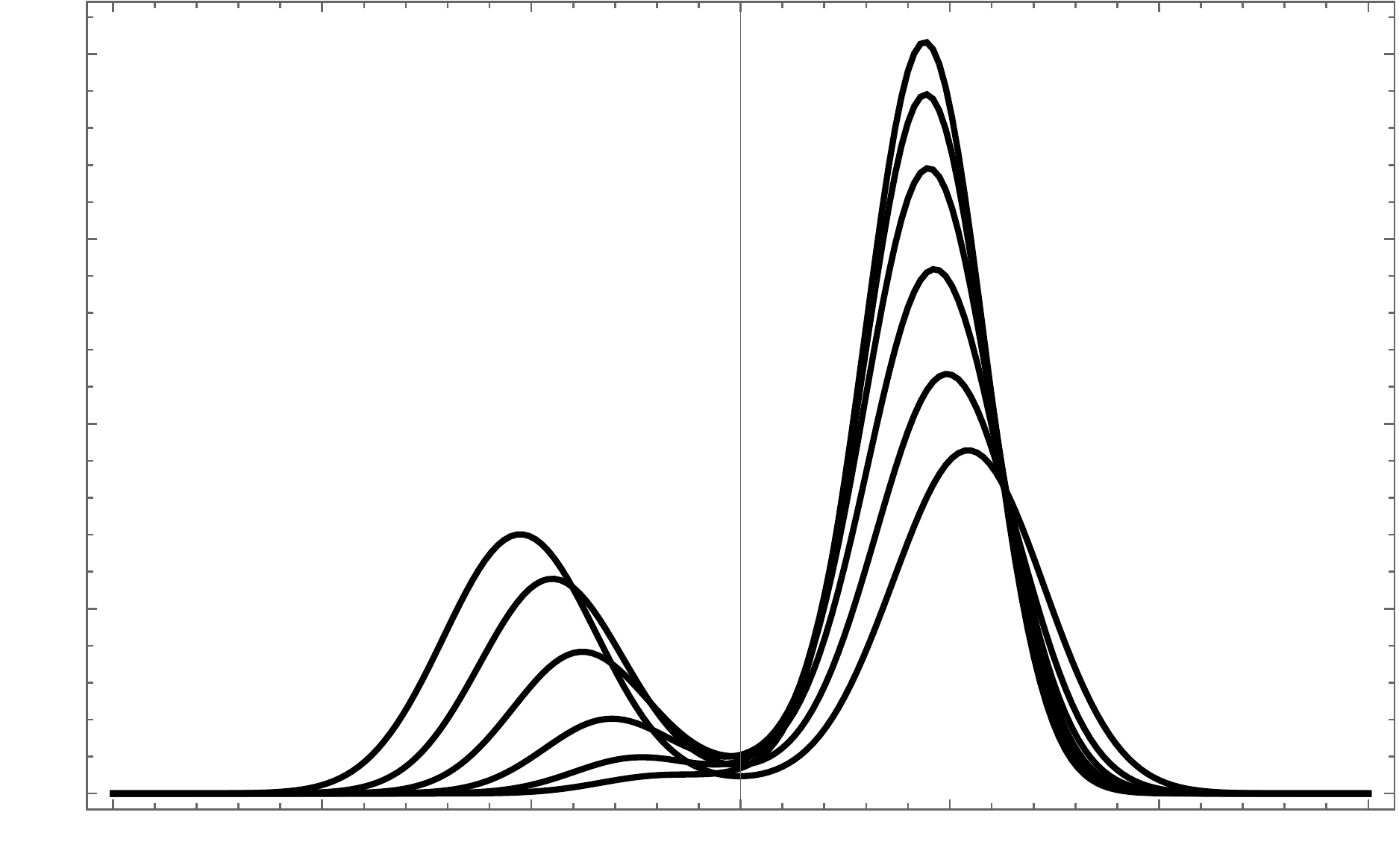}{fig5}
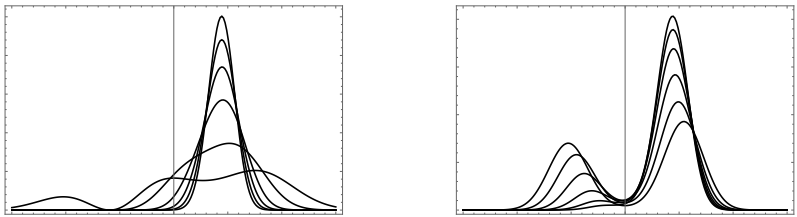
\begin{figure}[h]
\begin{center}
\begin{tikzpicture}[scale=0.6]
\pgftext[at=\pgfpoint{0cm}{0cm},left,base]{\pgfuseimage{fig4}} 
\pgftext[at=\pgfpoint{10cm}{0cm},left,base]{\pgfuseimage{fig5}} 
\end{tikzpicture}
\caption{\footnotesize  Two examples of the localization of a wave function. In both case, the wave packet localizes around a coherent state with an average position and momentum that matches the initial average position and momentum.}
\label{f4} 
\end{center}
\end{figure}

The solutions of the effective Schr\"odinger equation defined by $H_\text{eff}$ in \eqref{hef}---the deterministic part of the unravelling shared by QSD---have some characteristic properties. The non-Hermitian part of the Hamiltonian has the tendency to localize wave packets towards states that are annihilated by the branch creation operators
\EQ{
J_i\ket{\pi}=0\ .
\label{euu}
}
If the conditioned state were precisely $\ket{\pi}$ then the rate $r_i=\VEV{J_i^\dagger J_i}_\psi$  would vanish \eqref{ckk}. This tendency to localize will be counteracted by the opposite tendency for the wave function to spread out. This is unavoidable, even in a free system, but is particularly marked in a chaotic system.
For the particle with a single Lindblad operator \eqref{mnm}, the localized states \eqref{euu} are simply coherent states $A\ket{\pi}=a\ket{\pi}$ which have a have a spatial spread $\hbar/\sqrt{mkT}$, the de Broglie thermal wavelength. The localization effect is illustrated in figure \ref{f4}. 

The expectation values $\bar x=\VEV{x}$ and $\bar p=\VEV{p}$ now give Newton's Equations because in contradistinction to \eqref{rvv} the state is microscopically localized and Ehrenfest's Theorem now applies $\VEV{V'(x)}_\psi\approx V'(\bar x)$ and so
\EQ{
\frac{d\bar x}{dt}=\frac{\bar p}m\ ,\qquad \frac{d\bar p}{dt}=-V'(\bar x)-2\gamma\bar p\ .
}

However, it may not be reasonable to expect that the conditioned state becomes exactly localized on a coherent state due to the opposite effect driven by the non-linearity of the force $V'(x)$ and a more careful analysis is required. The effective Hamiltonian takes the form
\EQ{
H_\text{eff}=\frac{p^2}{2m}+V(x)+\frac\gamma2(xp+px)+\frac{i\hbar}2\big(\VEV{A^\dagger}_\psi J-\VEV{A}_\psi J^\dagger\big)-\frac{i\hbar}2J^\dagger J\ .
}
It is the non-Hermitian term 
\EQ{
J^\dagger J=\frac{4\gamma mkT}{\hbar^2}(x-\VEV{x}_\psi)^2+\frac\gamma{4mkT}(p-\VEV{p}_\psi)^2+\gamma\ ,
}
that drives the localization of the state towards a coherent state. We can quantify this in terms of the time it takes to localize the state over a spatial length $\ell$:
\EQ{
\tau\thicksim \frac{\hbar^2}{\gamma m kT}\cdot\frac1{\ell^2}\ .
}
In a free theory, the wave packet would become a coherent state. However, the non-linearities of the potential have the opposite effect of making the wave packet spread out. The most extreme example of this happens when the associated classical system is chaotic. In this case, we can estimate this spreading because it is known to mirror the exponential divergence of nearby classical trajectories: $\delta x(t)\sim\delta x(0)\exp(\lambda t)$, where $\lambda$ is an effective Lyapunov exponent. This latter quantity will depend on the point in phase space over which the wave packet sits. The localizing and spreading will balance when $\tau\sim\lambda^{-1}$, so that the size of the wave packet at some point in phase space will be roughly\footnote{This length scale is the same as the estimate of the coherence length of the unconditioned state \cite{PZ}.}
\EQ{
\ell\thicksim\hbar\cdot\sqrt{\frac{\lambda}{\gamma m kT}}\ ,
\label{zsz}
}
so larger by a factor $\sqrt{\lambda/\gamma}$ than the coherent state. However, the scale is still minute on macroscopic scales, so, even in a chaotic system, the conditioned state will be localized sufficiently well  that Ehrenfest's theorem applies and Newton's equations are valid for the centre of the wave packet. However, as the particle mass is reduced we expect it to reach a regime where it classically undergoes Brownian motion. Can we see this from the dynamics of the conditioned state? 

\vspace{0.3cm}
\noindent{\bf Quantum jumps.}~~~So far we have not discussed the jumps that can occur on top of the non-linear deterministic dynamics defined $H_\text{eff}$ whenever the unconditioned state generates a new branch. To start with let us estimate the its rate. If we assume that the wave packet is defined by a single length scale $\ell$ as in \eqref{zsz}, then we have
\EQ{
r\thicksim \frac{\gamma m kT}{\hbar^2}\ell^2\thicksim \lambda\ .
\label{iyy}
}
So the effective Lyapunov exponent determines the rate of the jumps in a system that is classically chaotic.

\pgfdeclareimage[interpolate=true,width=6cm]{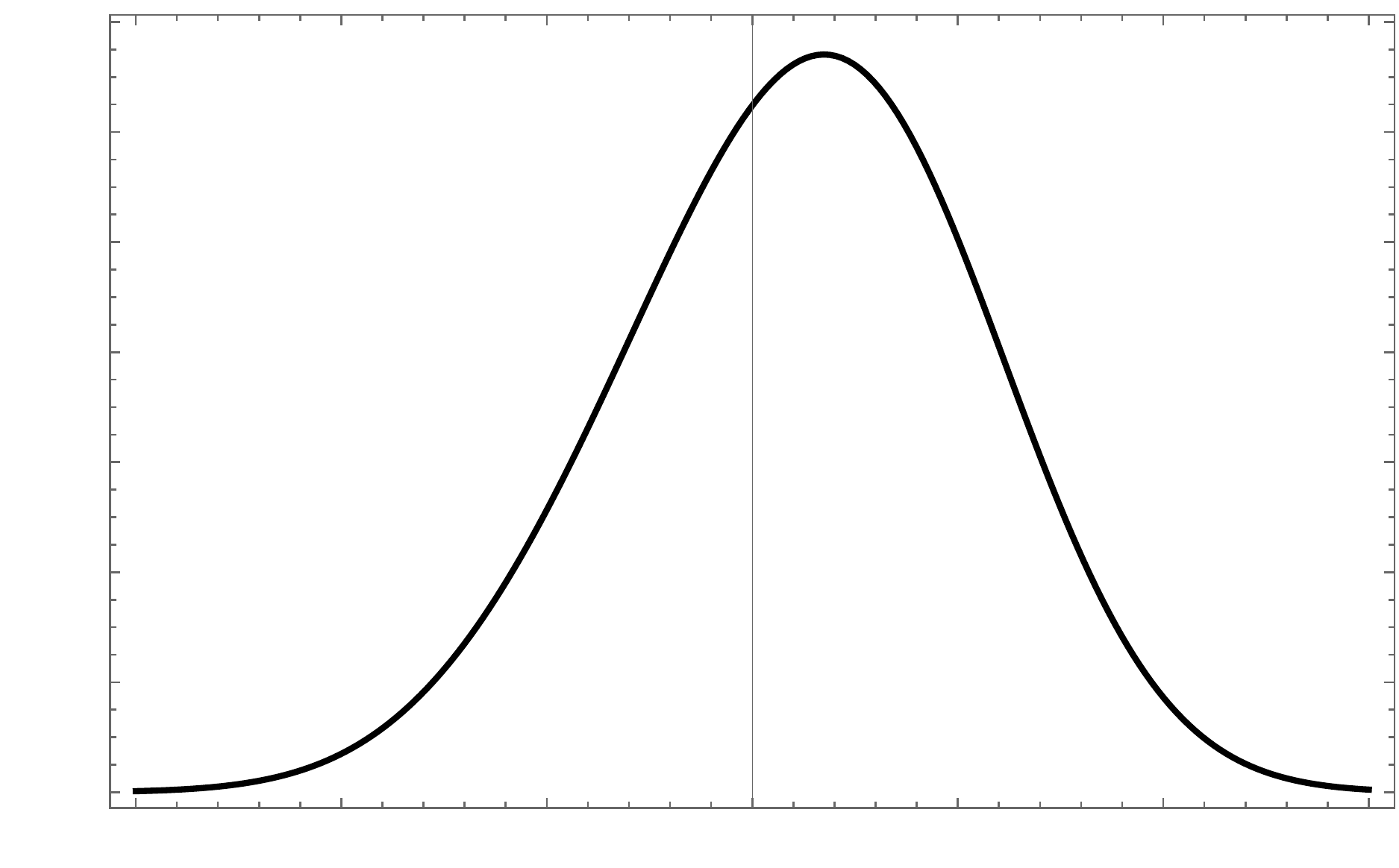}{fig6}
\pgfdeclareimage[interpolate=true,width=6cm]{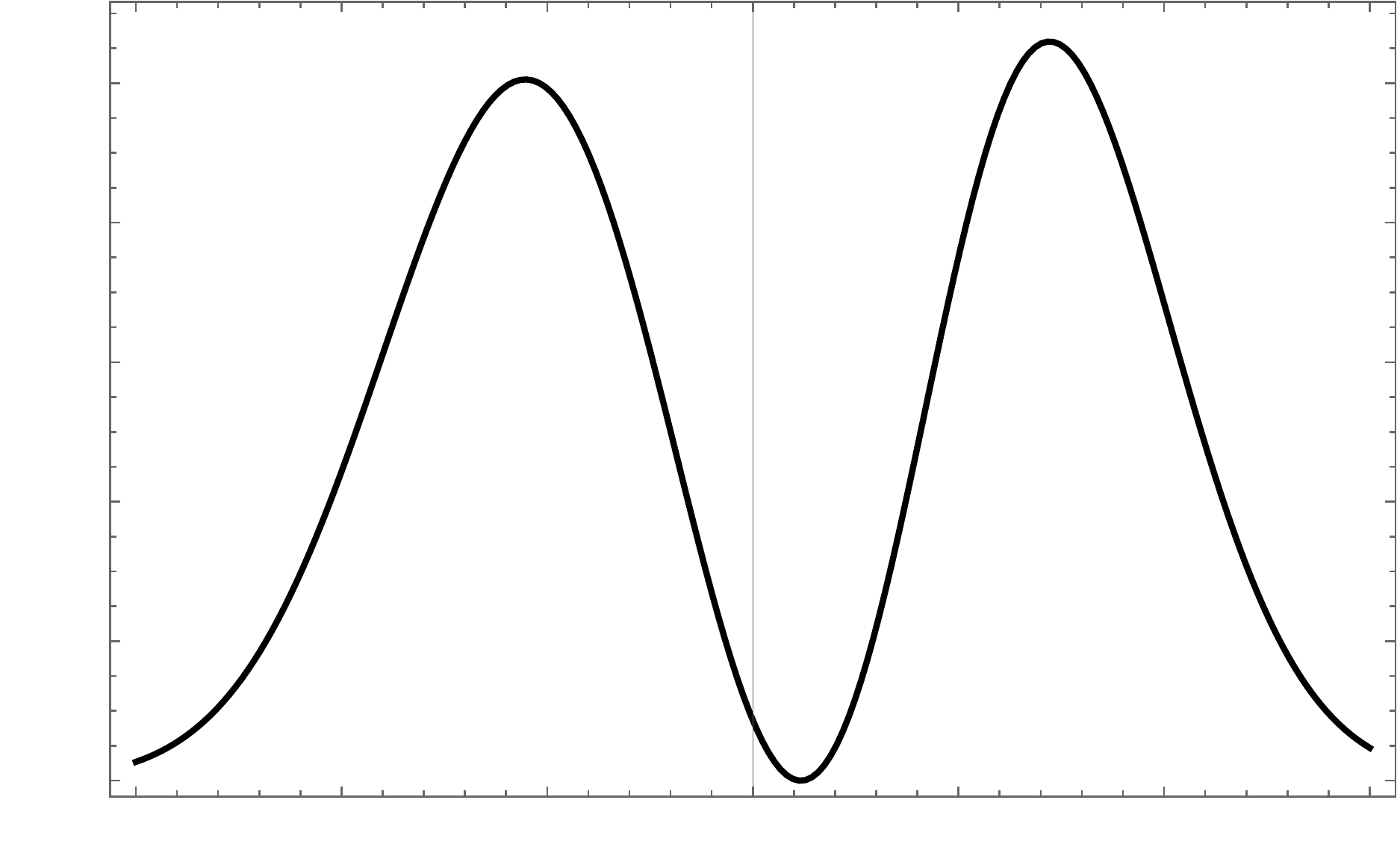}{fig7}
\pgfdeclareimage[interpolate=true,width=6cm]{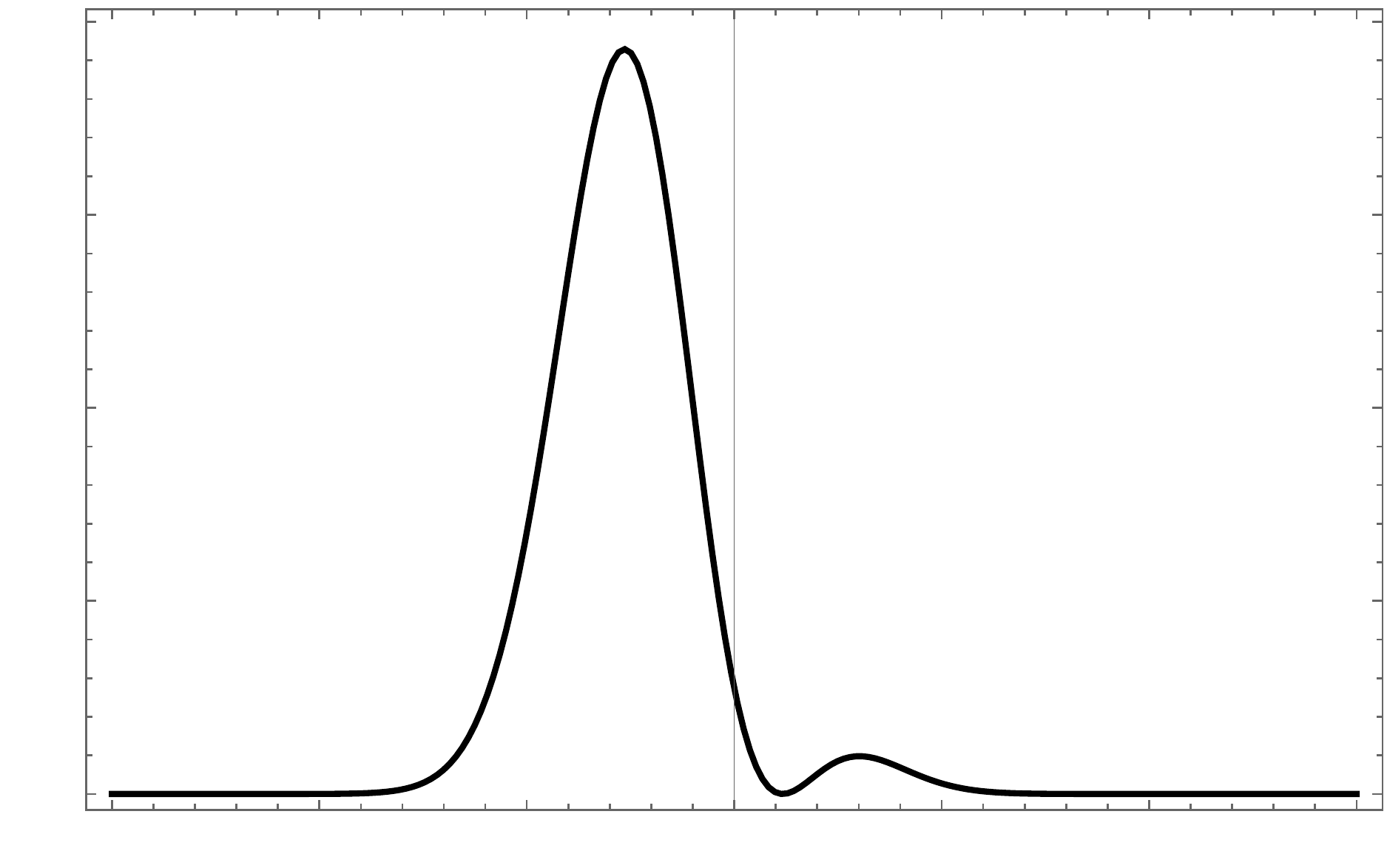}{fig8}
\begin{figure}[h]
\begin{center}
\begin{tikzpicture}[scale=0.65]
\pgftext[at=\pgfpoint{0cm}{6cm},left,base]{\pgfuseimage{fig6}} 
\pgftext[at=\pgfpoint{7cm}{6cm},left,base]{\pgfuseimage{fig7}} 
\pgftext[at=\pgfpoint{14cm}{6cm},left,base]{\pgfuseimage{fig8}} 
\node at (6.75,8.5) {$J$};
\draw[very thick,->] (6.1,7.4) -- (7.2,7.4);
\draw[very thick,->] (13.1,7.4) -- (14.2,7.4);
\node at (13.65,8.5) {\small $H_\text{eff}$};
\end{tikzpicture}
\caption{\footnotesize  The probability density of a wave packet undergoing a jump. In the subsequent non-linear dynamics dictated by $H_\text{eff}$, one of the pair of resulting wave packets survives by gaining all the probability.}
\label{f4} 
\end{center}
\end{figure}
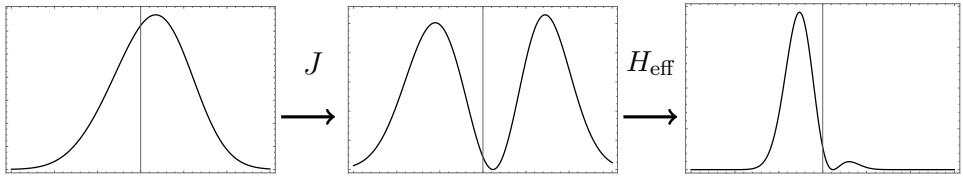

The phenomenology of the jumps in a free theory has been considered in \cite{SH} (see also \cite{BH1,BH2}). At a detailed level, the action of $J$ has the effect of splitting a wave packet into two separate wave packets. The non-linear dynamics then takes over and one of the offspring is amplified while the other fades away over a time scale $\lambda^{-1}$, the average time between jumps. The one that survives depends on the detailed form of the initial wave packet. Effectively a jump shifts the original wave packet by a small amount sideways. 

This rather simple phenomenology will allow us, following \cite{SH}, to simplify what seems to be a complicated problem arising from the fact that the exact dynamics depends on the detailed form the wave packet. Effectively, at the coarse grained level, we can model the dynamics as if the wave packet has, on average, a fixed shape determined solely by the length scale $\ell$ but with a position in phase space that varies stochastically. The stochastic increment of the position of the wave packet follows from \eqref{uru}
\EQ{
d\bar x=\frac{\bar p}m\,dt+\gamma_x\,\big(dN-r\,dt\big)\ ,\qquad d\bar p=-V'(\bar x)\,dt-2\gamma\bar p\,dt+\gamma_p\,\big(dN-r\,dt\big)\ ,
\label{lang}
}
where
\EQ{
\gamma_x=\frac{\VEV{J^\dagger(x-\bar x)J}_\psi}{\VEV{J^\dagger J}_\psi}\ ,\qquad \gamma_p=\frac{\VEV{J^\dagger(p-\bar p)J}_\psi}{\VEV{J^\dagger J}_\psi}\ .
\label{dd1}
}
In the above, the conditioned state is localized on macroscopic scales and so we have exploited Ehrenfest's Theorem to approximate $\langle V(x)\rangle\approx V(\bar x)$. 
Apart from this approximation, these are exact equations where the quantities $\gamma_x$ and $\gamma_p$ depend on the detailed form of the wave packet and satisfy stochastic evolution equations themselves. 
What we are after is an effective coarse grained description valid at macroscopic phase space and time scales.
In this description, we ignore the dynamics of the detailed form of the wave packet and so replace $\gamma_{x,p}\to\bar\gamma_{x,p}$ by non-stochastic time-averaged quantities.
In addition, at macroscopic scales many jumps occur and the Poisson process looks like a continuous random walk, or Wiener process. The appropriate limit is
\EQ{
dN\longrightarrow r\,dt+\sqrt r\,dW\ .
}
Applying this approximation, the stochastic increments \eqref{lang} (in the semi-classical limit) become the Langevin equations
\EQ{
\frac{d\bar x}{dt}=\frac{\bar p}m+\sqrt{\bar r}\bar\gamma_x\,\xi\ ,\qquad \frac{d\bar p}{dt}=-V'(\bar x)-2\gamma\bar p+\sqrt{\bar r}\bar\gamma_p\,\xi\ ,
\label{lang}
}
where $\xi(t)=dW/dt$ is Gaussian noise with stochastic correlators
\EQ{
\mathscr E\big(\xi(t)\big)=0\ ,\qquad \mathscr E\big(\xi(t)\xi(t')\big)=\delta(t-t')\ .
}
In \eqref{lang}, we have also coarse grained the rate $r\to\bar r$.
We can estimate the time averaged quantities in terms of the wave packet scale $\ell$:
\EQ{
\bar r\bar\gamma_x^2\thicksim\frac{\hbar^2\lambda^2}{\gamma mkT}\ ,\qquad \bar r\bar\gamma_p^2\thicksim\gamma m kT\ .
\label{est}
}

Finally, we can see whether this effective description is consistent, by considering the Fokker-Planck equation that describes the evolution of the probability density function $P(x,p)$ for a ensemble of classical trajectories described by the Langevin equations \eqref{lang}:
\EQ{
\frac{\partial P}{\partial t}=\{H,P\}+2\gamma p\frac{\partial P}{\partial p}+\frac{\bar r\bar\gamma_p^2}2\frac{\partial^2P}{\partial p^2}\ .
\label{fpe}
}
Since, at the coarse grained level, the conditioned state is effectively point-like in phase space and the unconditioned state is very decoherent (e.g.~has large entanglement entropy), the Wigner function of the unconditioned state 
\EQ{
{\cal W}(x,p)=\int_{-\infty}^\infty\frac{dx'}{2\pi}\,\rho(x+x'/2,x-x'/2)e^{ix'p}\ ,
}
acts as a probability density for the conditioned state in phase space. This means that we can identify $P$ with ${\cal W}$. Indeed, it is known, that in the semi-classical limit when the conditioned state is decoherent, the master equation \eqref{mst} written in terms of the Wigner function of the unconditioned state is precisely of the form \eqref{fpe} \cite{PZ,Zurek2}. This identifies the coarse grained quantity $\bar r\bar\gamma_p^2=4\gamma mkT$, a constant (note the consistency with the estimate \eqref{est}). The Langevin equations \eqref{lang} are then
\EQ{
\frac{d\bar x}{dt}=\frac{\bar p}m\ ,\qquad \frac{d\bar p}{dt}=-V'(\bar x)-2\gamma\bar p+\sqrt{4\gamma mkT}\,\xi\ ,
\label{lang2}
}
precisely the standard equations of classical Brownian motion. 

The conclusion is that the coarse grained dynamics of the wave packet on scales which do not resolve the detailed form of the wave packet or the individual jumps will be described by the Langevin equation of classical Brownian motion. But if we do not coarse grain, then we can draw the very satisfying conclusion that, in the semi-classical limit, the jumps of the wave packet become the analogues of the individual kicks of classical Brownian motion.

Finally on a historical note, it is interesting to note that early in the history of quantum mechanics it was hoped that particles would correspond to localized solutions of Schr\"odinger's equation \cite{Sch2}. It was soon realized that this was untenable because solutions do not remain localized. Now we see that it is precisely an effective, stochastic, Schr\"odinger equation \eqref{hee} that has the original desired property and that particle-like states only exist in the subsystem frames on account of their pervasive entanglement with the environment.

\vspace{0.2cm}
This work was supported in part by STFC grant ST/P00055X/1.

\end{document}